\documentclass[10pt,conference]{IEEEtran}
\usepackage{graphicx}
\usepackage{enumerate} %
\usepackage{multirow}
\usepackage[small,belowskip=-12pt]{caption}
\usepackage{subfig}
\usepackage{epsfig}
\usepackage{mdwlist}
\usepackage{array}
\usepackage{setspace}

\usepackage{cite}

\ifCLASSINFOpdf
\else
\fi
\usepackage[cmex10]{amsmath}
\begin{document}
\IEEEoverridecommandlockouts
\newtheorem{definition}{Definition}
\newtheorem{lemma}{Lemma}
\title{Coding the Beams: Improving Beamforming Training in mmWave
Communication System}

 \author{\IEEEauthorblockN{Y. Ming Tsang, Ada S. Y. Poon}
 \IEEEauthorblockA{Department of Electrical Engineering, Stanford University\\
  ymtsang@stanford.edu, adapoon@stanford.edu} \and
 \IEEEauthorblockN{Sateesh Addepalli}
 \IEEEauthorblockA{Cisco Systems Inc\\
  sateeshk@cisco.com}
\thanks{This work is supported by Croucher Foundation Fellowship and Cisco Systems Inc.}}
\maketitle

\begin{abstract}
The mmWave communication system is operating at a regime with high number of antennas and very limited number of RF analog chains. Large number of antennas are used to extend the communication range for recovering the high path loss while fewer RF analog chains are designed to reduce transmit and processing power and hardware complexity. In this regime, typical MIMO algorithms are not applicable.

Before any communication starts, devices are needed to align their beam pointing angles towards each other. An efficient searching protocol to obtain the best beam angle pair is therefore needed. It is
called BeamForming (BF) training protocol.

This paper presents a new BF training technique
called beam coding. Each beam angle is assigned unique signature code. By
coding multiple beam angles and steering at their angles simultaneously in a
training packet, the best beam angle pair can be obtained in a few packets. The proposed BF training technique not only shows the robustness in non-line-of-sight environment, but also provides very flat power variations within a packet in contrast to the IEEE 802.11ad standard whose scheme may lead to large dynamic range of signals due to beam angles varying across a training packet.
\end{abstract}
\IEEEpeerreviewmaketitle

\section{Introduction}
Millimeter Wavelength (mmWave) communication system has been
considered as a viable candidate to realize multiple Gigabits per second (Gbps)
data rate. A 7-GHz
of bandwidth at 60~GHz carrier frequency is allocated by
most regulatory bodies worldwide for unlicensed
uses~\cite{yong07}. The rather short wavelength allows more antennas to be
deployed in miniature consumer devices. Consequently, IEEE standards, namely, 802.15.3c\cite{IEEE802153c} and 802.11ad\cite{IEEE80211ad}, are in active development to standardize the protocols and requirements for mmWave system.

One of the problems for communicating at the mmWave spectrum is its
high signal path loss. By Friis free space formula, signals at 60 GHz suffer from 28 dB
more loss than that at the 2.4 GHz band. A large number of antennas
are therefore placed at the transceiver to exploit the beamforming gain for extending the communication range.
For example, 16 antennas are supported in the
standards and in some mmWave Radio-Frequency Integrated Circuits (RFICs) \cite{garcia10}\cite{emanuel10}.

Ideally, antennas should connect to the baseband processor via separate RF/analog downconversion chains (or simply called analog chains) so that signal processing at spatial domain can be executed at digital
baseband. However, multiple analog chains are costly at mmWave frequency and sampling the analog signal at GHz
rate consumes substantial amount of power. Hence, only one analog chain is found in current mmWave RFIC design \cite{garcia10}\cite{emanuel10}.

In this regime of high number of antennas and very few analog chains, typical MIMO system techniques are not applicable. Antenna selection \cite{molisch04} was proposed for this regime but it only provides limited transmission range extension. To generate beamforming effect for the range extension, all antennas are active during
communications. The weight at each antenna is designed in such a way that the signals from all antennas are coherent at the desired directions. Moreover, the weights are applied at the RF \cite{garcia10} so that only one analog chain is needed to create highly directional beam patterns.

The directional beam pattern improves the transmission range but it complicates communication protocol designs. Communications between two devices are not possible if their beam directions are not pointing towards each other. Therefore, an efficient protocol that discovers the best beam direction pair between devices is very crucial. This protocol is called BeamForming (BF) training. In this paper, the best beam direction pair refers to the beam direction pair that gives the highest channel gain. Also, we focus on the BF training between the Access Point (AP) and devices. But the same protocol can be applied to a device-to-device link setup and other metrics that define the best beam direction pair.

The BF training process begins with the transmitter sending training packets at a set of pre-defined beam directions (or simply called beams). The number of beam depends on the number of antennas and the amount of coverage \cite{maltsec10} required. For example, for Uniform Linear Array (ULA) with 16 antennas, 32 beams are suggested. One method is to exhaustively examine all the beam pairs \cite{Dai06efficientbroadcasting} by sending a training packet for each beam pair. However, the searching time is prohibitively long. To reduce the training time, Wang
\cite{Wang09} and the standards \cite{IEEE802153c}\cite{IEEE80211ad} propose multi-level training scheme. The search
starts from Lower-Resolution (L-Re) beams which cover larger space per beam.
In the next level, higher-resolution (H-Re) beams covered by the selected
L-Re beams in earlier stage are searched. While the training time is greatly reduced, multi-level scheme is shown incompetent in identifying non-LOS (NLOS) paths as the multiple path resolubility \cite{Tse05} is poor in the L-Re beams in earlier stage. Since an individual packet is sent per beam pair in the above schemes, they are categorized as Packet-by-Packet (PbP) BF training.

A training packet is required to be sent in PbP training for training a beam pair. This introduces high overheads since headers and preamble in a packet provide no information for BF training but they are used only for message identification purposes. In IEEE 802.11ad, in-packet BF training packet structure is purposed to reduce the overheads in the beam refinement step which executes H-Re BF training in multi-level scheme. It allows varying beams within a training packet. Hence, training multiple beams only requires the amount of network bandwidth equivalent to one packet transmission.

Despite the advantages of using In-packet training, it poses some challenges in the implementation. First, the changes of beam directions significantly increase the dynamic range of the signals across a packet. Resetting Automatic Gain Control (AGC) may be needed for every change of beam direction. Second, even if the AGC is reset for every change of beam direction, the timing synchronization point may be altered after each AGC re-settling. The relative delays and phase differences among all beam direction pairs may become inaccurate \cite{kasher10} (comment 414). Consequently, in the IEEE 802.11ad standard, the change of AGC gain preceding each beam direction change is disallowed so that the relative delay information is preserved. To limit the increase of dynamic range, in-packet packet structure is only allowed during beam refinement step in which only a smaller set of beams are trained.

If the power variations of signal during the BF training in an in-packet training packet is reasonably flat, it allows in-packet training to be implemented in every stage of BF training to unleash the full potential of the in-packet training. The question is how one can provide relatively flat power variations while employing in-packet training.

As it can be observed, receive power variations in PbP training are very limited as the beam direction is maintained the same throughout the packet. To enable this feature for in-packet training, we propose a beam coding scheme in which the array is steered at multiple highest-resolution beams simultaneously. Each beam is assigned an unique signature sequence for receivers to differentiate beam angles and extract the channel
information of all the beam pairs. Since the
covering beam pointing directions are the same throughout the packet, beam coding provides a fairly uniform receive power across a packet.

With only one analog chain in the transceiver path, how can beams carry different signature sequences? Instead of encoding the beams at the baseband processor, the antenna weights at the RF are manipulated as each antenna is weighted individually. Antenna weights in an array can be uniform or nonuniform \cite{balanisBook}\cite{treesBook}. Non-uniformly weighted array allows magnitude variations on weights while uniformly weighted array varies the phases of weights only. The beam coding scheme may require nonuniform weighted array for the best performance, which is supported by the
circuitry in \cite{tsang10}. In fact, we will show by using specific signature sequences, beam coding in uniformly and nonuniformly weighted arrays have similar performance.

\section{In-packet BF training}
\begin{figure}
\centering
\epsfig{file=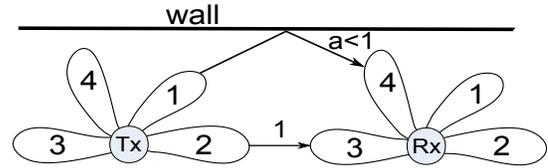,height=0.85in, width=2.8in}
\caption{shows a communication scenario: both devices are allowed
to steer at the 4 beams only for communication.}\label{fig:toy}
\end{figure}

In this section, the background of in-packet training is provided. The packet structure in IEEE 802.11ad for in-packet training is first described. Then we discuss two BF training algorithms that utilize the packet structure. For the sake of presentation, we restrict the transmitter and the receiver being able to steer their arrays at 4 beams specified in Fig. \ref{fig:toy}. It can be easily generalized to the setup with more beams. In the example, there are one LOS path and one NLOS path with attenuation $a<1$.

\begin{figure}
\centering
\epsfig{file=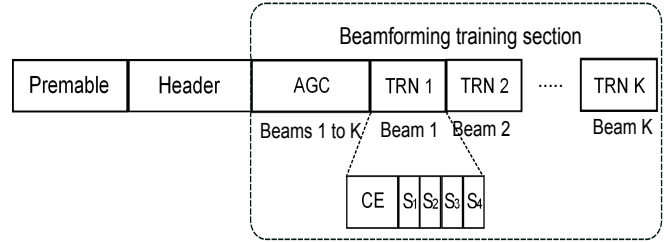,height=1.25in, width=3.4in}
\caption{shows the packet structure for in-packet training in IEEE 802.11ad: CE represents channel estimation sequences and $S_1$ through $S_4$ are subfields for estimating the phases and amplitudes of channel taps}\label{fig:struc80211ad}
\end{figure}

\subsection{IEEE 802.11ad packet structure for in-packet training}
IEEE 802.11ad provides the capability for in-packet BF training. The packet structure shown in Fig. \ref{fig:struc80211ad} begins with the preamble and header sections which are steered using L-Re beams similar to multi-level PbP training to cover larger space. Following the header section, the BF training section allows training in H-Re beams. It contains AGC field and training (TRN) fields. The set of beam directions that are being trained are changed sequentially within the AGC field for a receiver to calculate an appropriate AGC gain. Then, a TRN field is allocated for each beam for channel estimation. It contains Channel Estimation(CE) sequence and extra subfields to improve the accuracy of tap delay estimation. It is worth noting that even though beams are changed every TRN field, the AGC gain is fixed across all the TRN fields. Hence, larger dynamic range of signals can be expected.

\subsection{Exhaustive in-packet training}
The exhaustive search becomes feasible as the overheads in header and preamble sections are minimized through the use of in-packet training. The number of training packets in the example (Fig. \ref{fig:toy}) is reduced from 16 to 4.

\begin{figure}
\centering
\epsfig{file=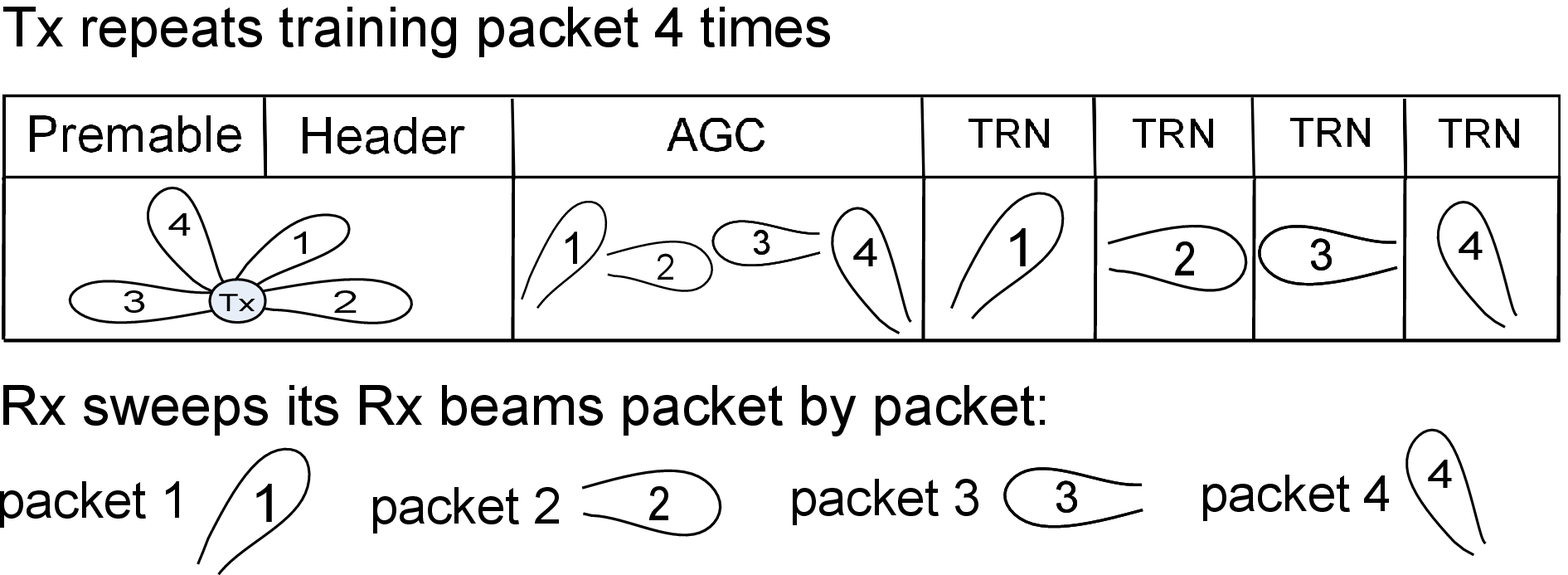,height=1in, width=3.2in}
\vspace{-6pt}
\caption{shows the flow for exhaustive in-packet training}\label{fig:exhaustive_inpacket}
\vspace{+4pt}
\end{figure}

The exhaustive in-packet training procedures are shown in Fig
\ref{fig:exhaustive_inpacket}. The transmitter sends the same training
packet 4 times while the receiver steers its beams one by one. Since beam pairs
(2, 3) and (1, 4) are both aligned and $a<1$, the receiver is able to identify
(2, 3) as the best beam pair. It is also worth noting that since the receiver is
able to obtain full channel information about all the combinations of Tx and Rx
highest-resolution beams, this scheme has compelling performance in NLOS environment.

\subsection{Feedback In-packet training} \label{sec:feedback}
The BF training procedure described previously requires no explicit feedback from the
receiver until the end of training. In fact, limited explicit feedback consumes very few resources. For example, the association request that is sent for receivers to associate with the AP can be used to piggyback the intermediate training results to improve the training time.

\begin{figure}
\centering
\epsfig{file=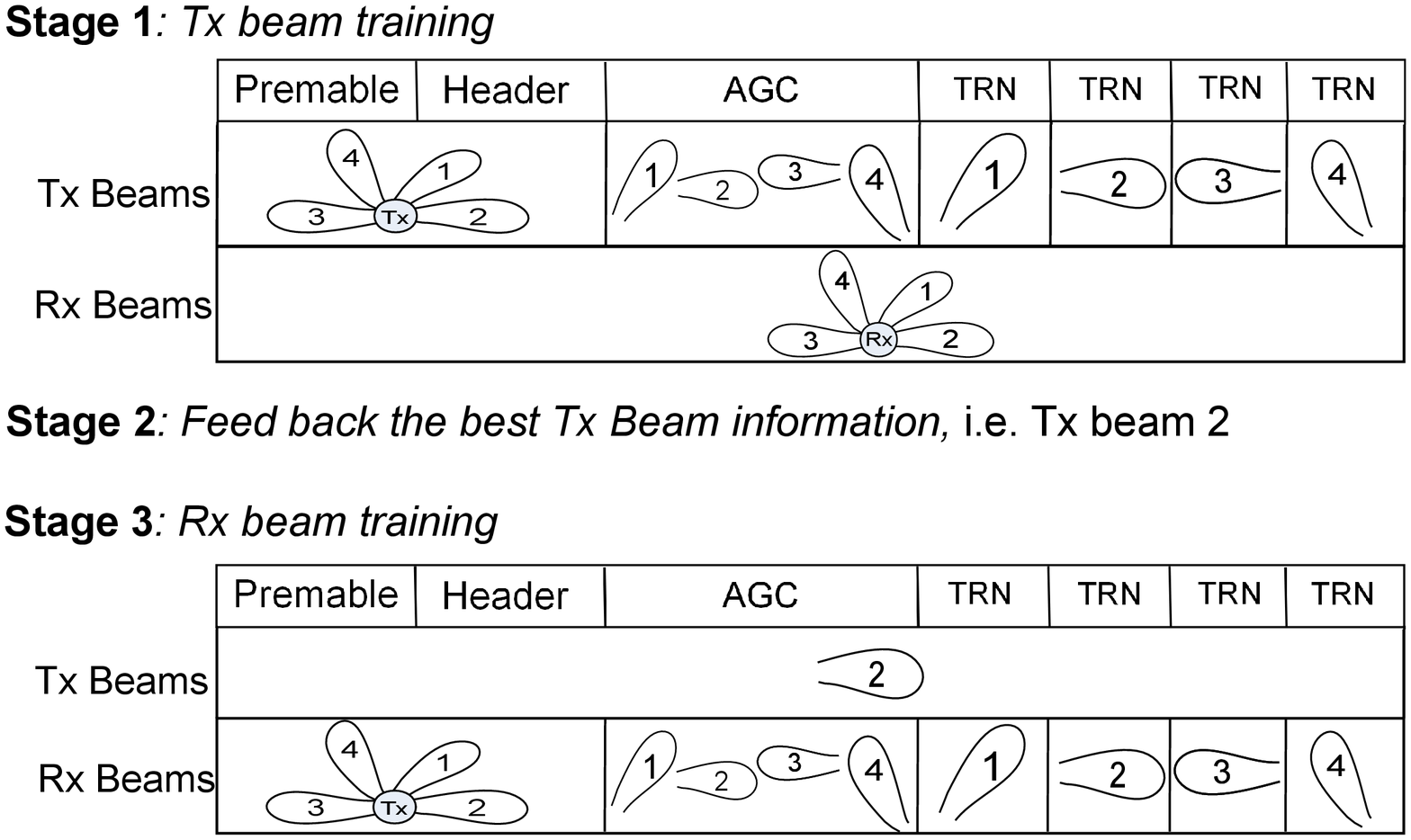,height=2in, width=3.4in}
\caption{shows the flow for feedback in-packet training.}\label{fig:in_packet_feedback}
\end{figure}

The procedure is shown in Fig. \ref{fig:in_packet_feedback}. The Tx beam is first being trained. The transmitter forms its training packet similar to the exhaustive in-packet training while the receiver is steering at all its directions simultaneously. The received signal is $y = 0.5[a\;\;1\;\;0\;\;0]$, where $y[t]$ represents the receive signal amplitude at the TRN field $t$. Since $a<1$, Tx beam 2 is chosen. Then, this information is fed back to the transmitter which then sends a training packet through its beam 2 at stage 3. Receiver is now varying its beam directions within the packet and detects Rx beam 3 as the best Rx beam. Hence the training can be completed in 2 training packets.

Unlike exhaustive in-packet training, the feedback scheme does not provide full channel information since only Tx beams are trained at stage 1. Moreover, since the receive BF gain is lowered by having all Rx beams active at stage 1, the communication range is reduced. This poses an interesting tradeoff between the training time and the range, which cannot be leveraged easily in PbP training. The tradeoff is explored in our future work.


\section{Beam coding}
Employing in-packet BF training may result in larger dynamic range across a packet. How can one provide reasonably flat power variations across a packet?

Instead of varying the beam pointing directions in each TRN field as in the standard, if a scheme allows multiple highest-resolution beams being steered at the same time across all the TRN fields, the receive power variations are greatly reduced and no explicit AGC resettling is needed. The
main barrier for such protocol is the single analog chain that forces signals at each beam to be identical. Two transmit beams are
therefore indistinguishable at the receiver if they are steered at the same time. To shed some light on problem, it can be observed that the weight of each
antenna can be adjusted individually by tuning a phase-shifter on each antenna. For instance, if an additional phase $\pi$ is added to the weight
at the $n^\text{th}$ antenna, all signals through the $n^\text{th}$ antenna
will be negated. We leverage this observation and propose a scheme called \emph{beam coding}.

Beam coding first steers the antenna array towards multiple beam direction as described in section \ref{sec:sametime}. By encoding each beam with a signature sequence, a receiver is able to identify the signal strength from each beam direction as illustrated in section \ref{sec:beam_coding_tx}. To further reduce the power variations, the beam directions chosen for training within a training packet are designed to be orthogonal as discussed in section \ref{sec:orthogonal}.
For the sake of presentation, we assume that all antenna elements are
omnidirectional and the array is a non-uniformly weighted linear array with constant
distance between neighboring antennas. The normalized
inter-antenna distance
is denoted as $\Delta_d=\frac{d}{\lambda}$ where $d$ is the inter-antenna distance
and $\lambda$ is
the wavelength of the signal. We will also show that applying beam coding in uniform weighted array is also feasible.

\subsection{Transmit at two beams at the same time} \label{sec:sametime}
Assume the system with $N$ antennas, if the transmit data consists of all ones,
the signal at angle $\phi$ is
\begin{small}
\begin{align}
x(\phi)=\sum_{n=0}^{N-1} w_n e^{j2\pi n \Delta_d \cos\phi}
\label{eq:transmitBeam}
\end{align}
\end{small}
where $w_n$ is the weight at the $n^{\text{th}}$ antenna. If we want to steer
the angle to the transmit beam direction $\phi_1$, the antenna weight will
become $w_n=e^{-j2\pi n \Delta_d \cos\phi_1}$.

Now, to transmit signal through two beam directions, $\phi_1$ and $\phi_2$, at
the same time, the antenna weights are
\begin{small}
\begin{equation}
w_n=e^{-2\pi n \Delta_d \cos \phi_1} + e^{-2\pi n \Delta_d \cos \phi_2}
\label{eq:superimpose}
\end{equation}
\end{small}
The resulting transmit signal equals to
\begin{small}
\begin{align*}
x(\phi)=\sum_{n=0}^{N-1} e^{-2\pi n \Delta_d(\cos
\phi_1-\cos\phi)}+\sum_{n=0}^{N-1}
e^{-2\pi n \Delta_d(\cos \phi_2-\cos\phi)}
\end{align*}
\end{small}
This implies that due to the linearity of antenna weights in the transmit
signal, the weights for the two beams are superimposed and create a new beam pattern which attains the maximum points at $\phi=\phi_1$ and
$\phi=\phi_2$. Antenna weights for 4 transmit beams are formed similarly. In analogy, the antenna weights can be formed at the receiver side for
4 beams.

\subsection{Encoding the beams} \label{sec:beam_coding_tx}
\begin{figure}
 \centering
 \epsfig{file=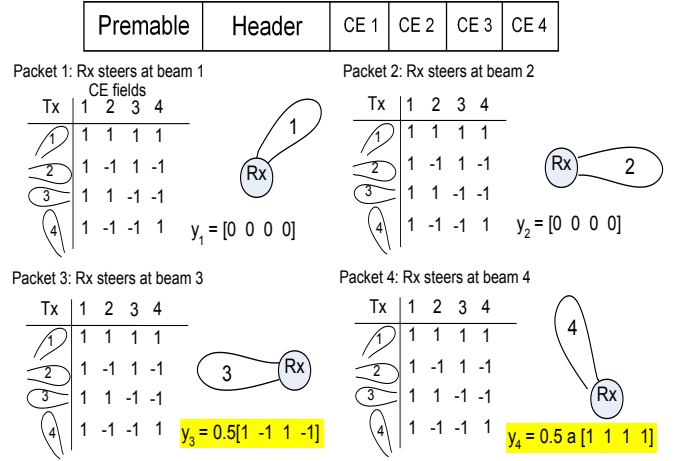,height=2.4in, width=3.4in}
 \caption{shows exhaustive beam coding training procedure.}\label{fig:beam_coding_full}
 \end{figure}

First, each transmit beam $p$ is assigned an unique signature sequence, $s^{Tx}_p$,
for all $p \in \{1,2,3,4\}$ as follows:
\begin{small}
\begin{align*}
&s^{Tx}_1 = [1 \;\;\;\;\;\;1 \;\;\;\;\;\;1 \;\;\;\;\;\;1] &s^{Tx}_2 = [1
\;\;-1 \;\;\;\;\;\;1\;\;-1]\\
&s^{Tx}_3 = [1 \;\;\;\;\;\;1 \;\;-1\;\;-1]
&s^{Tx}_4 = [1 \;\;-1 \;\;-1\;\;\;\;\;\;1]
\end{align*}
\end{small}
These signature sequences are derived from Walsh code and are known to both
transmitter and
receiver. Let $\beta_n(\phi)=e^{-2\pi n \Delta_d\cos \phi}$. Then, the antenna weights, denoted as $w^{Tx}_n[t]$ for all $t \in \{1,2,3,4\}$, are calculated for the consecutive CE fields in the TRN fields as shown in Fig. \ref{fig:beam_coding_full}, where $t$ represents the CE field index:
\begin{small}
\begin{align*}
w^{Tx}_n&=\frac{1}{\sqrt{4}}\left(s^{Tx}_1 \beta_n(\phi_1)+ s^{Tx}_2
\beta_n(\phi_2)+s^{Tx}_3
\beta_n(\phi_3)+ s^{Tx}_4 \beta_n(\phi_4)\right)
\end{align*}
\end{small}
Hence, the antenna weights at CE field indexes 1,2,3,4 are:
\begin{small}
\begin{align}
\text{At CE 1: }  w^{Tx}
_n[1]=0.5
\left(\beta_n(\phi_1)+\beta_n(\phi_2)+\beta_n(\phi_3)+\beta_n(\phi_4)\right)\nonumber \\
\text{At CE 2: }  w^{Tx}
_n[2]=0.5
\left(\beta_n(\phi_1)-\beta_n(\phi_2)+\beta_n(\phi_3)-\beta_n(\phi_4)\right)\nonumber\\%\label{eq:ant_weight}
\text{At CE 3: }  w^{Tx}
_n[3]=0.5
\left(\beta_n(\phi_1)+\beta_n(\phi_2)-\beta_n(\phi_3)-\beta_n(\phi_4)\right)\nonumber\\
\text{At CE 4: }  w^{Tx}
_n[4]=0.5
\left(\beta_n(\phi_1)-\beta_n(\phi_2)-\beta_n(\phi_3)+\beta_n(\phi_4)\right)\nonumber
\end{align}
\end{small}
Then, the transmitter is sending the same packet 4 times while
the receiver is switching beams after every packet duration.

While the receiver is steering at beam $q$, it receives $y_q[t]$. At the end of the
sweeping, it compares the received signal with the signature sequences by
calculating the correlations, $r_{(p,q)}=\sum_{t=1}^4 s^{Tx}_p[t]y_q[t]$.
Since beam pairs $(2,3)$ and $(1,4)$ are both aligned, we have $r_{(2,3)}=2$
and $r_{(1,4)}=2a$. Since $a<1$, the best beam pair $(2, 3)$ is obtained.

It takes 4 frames to complete the training. We call this as
\emph{exhaustive beam coding scheme}. The feedback scheme described in section \ref{sec:feedback} can also be leveraged via beam coding in the same way to further reduce the training time.

Walsh code are used to separate beams as its codewords are orthogonal to each other. Golay sequences possessing very good auto-correlation property are used as the CE sequences \cite{IEEE80211ad}. This property also helps protecting the Walsh code from losing orthogonality due to multi-path fading effect. Hence, receivers are able to identify the relative delays and signal strength in all the channel delay taps among all beam directions.

Apart from providing stable dynamic range of signals within a packet, beam coding also provides an essential structure for schemes such as compressed sensing\cite{candes08} to further reduce the training time. For instance, it can be done by replacing the signature sequences with Gaussian random sequences. However, since the sample size (i.e. the dimension of the antenna array) is rather small compared to other applications in compressed sensing, the reduction in training time is insignificant while the receiver complexity is much increased as computations are more involved. Hence, this benefit via beam coding is not further explored here.

\subsection{Orthogonal beam directions} \label{sec:orthogonal}
For each training packet, a set of beams are selected to be encoded by beam coding scheme. Since the scheme involves additions and subtractions on antenna weights
among the selected beams, different selections of beams have significant differences in power variations. To reduce power fluctuation, we show that the beams that are encoded in a packet are required to be orthogonal.

\begin{definition}
Let $\beta_n(\phi_i)$ and $\beta_n(\phi_j)$ be the antenna
weights for beams $i$ and $j$ at antenna $n$ respectively with $\sum_{n=0}^{N-1} |\beta_n(\phi_k)|^2 =1$, for $k=i, j$. Beams $i$ and
$j$ are orthogonal if and only if $\sum_{n=0}^{N-1}
\beta_n(\phi_i)\beta_n^*(\phi_j) = 0$, where $x^*$ is the complex conjugate of
$x$.
\end{definition}

To show that orthogonal beams reduce power fluctuation, the array that steers at beams $i$ and $j$ sets the antenna weights,
$w_n$, as $\frac{1}{\sqrt{2}}(\beta_n(\phi_i)\pm\beta_n(\phi_j))$, where the $\pm$
describes the signature sequences used for beam coding. Then,
\begin{small}
\begin{align*}
|w|^2= &\sum_{n=0}^{N-1} w_nw_n^* = \frac{1}{2}\sum_{n=0}^{N-1} |\beta_n(\phi_i)|^2 +
|\beta_n(\phi_j)|^2 \nonumber \\&\pm
\frac{1}{2}\left(\sum_{n=0}
^{N-1}\beta_n(\phi_i)\beta_n^*(\phi_j) + \beta_n(\phi_j)\beta_n^*(\phi_i)\right)
\end{align*}
\end{small}

If two beams are orthogonal, $|w|^2= 1$ and the transmission power is then
independent of the codes
used. Conversely, if the beams are
not orthogonal, the cross terms are non-zeros. In the extreme case
where $\beta_n(\phi_i) = \beta_n(\phi_j)$ for all $n$, the $|w|^2$ fluctuates
between 2 and 0 for different codes.

In fact, in a system with $N$ antennas, there are
at most $N$ orthogonal beams\cite{poon05}. The $N$ antenna weight variables create a N-dimensional space, resulting
in at most $N$ orthogonal vectors.

\subsection{Uniformly weighted phased array}
The beam coding scheme requires non-uniformly weighted
phased array for the best performance as the beam coding scheme varies the amplitudes across antenna weights. In fact, the scheme
also performs fairly well in uniformly weighted phased array in which only phases of the
weights can be adjusted. In
general, multiple beam steering creates larger
side lobes in uniform array. It can be showed that steering at two beams in
uniformly weighted Uniform Linear Array (ULA) introduces -9 dB side lobe level instead of -13 dB
in single beam steering. However, as shown in Fig. \ref{fig:polar_plot_2},
this is acceptable in the BF training because during BF training, all the beams are swept to
discover the best one for communication. The positions of side lobes will therefore be covered by the main lobes of other beams eventually.

In simultaneously training more than 2 beams, if the Walsh code is being used
as the signature sequences, and
the number of antennas is a power of 2, it can be proved analytically that beam coding is feasible in uniformly weighted ULA. The proof is omitted due
to the page limit. Instead, we will show by simulation that the performance in uniform array is very close to non-uniformly weighted array.

\section{Simulations}
In the simulation, the beam coding is compared with the scheme in the IEEE 802.11ad standard. The beam coding is demonstrated to provide comparably flat receive power variations across a packet.
We also show that beam coding has excellent performance even under uniformly weighted array.

Simulations follow the living room model in the
IEEE 802.11ad channel document
\cite{maltsec10} and the parameters are shown in Table \ref{tab:channel} and \ref{tab:simulation}.

\begin{table}
\centering
\begin{small}
\caption{Channel Model Parameters}
\vspace{-5pt}
\begin{tabular}{lcc} \hline
Path loss exponent & & 2 \\ \hline
\multirow{2}{*}{Cluster reflection loss} & (Mean, RMS) & (-10 dB, 4 dB) \\
& Truncation & -2 dB \\ \hline
\multirow{2}{*}{Intra-cluster model}& &Living room model \\
& & Section 5.5 in \cite{maltsec10}\\\hline
Intra-cluster AoA and & &\multirow{2}{*}{Gaussian (0, 5 deg)}  \\
AoD distribution & & \\\hline
\end{tabular}\label{tab:channel}
\end{small}
\end{table}
\begin{table}
\centering
\begin{small}
\vspace{+5pt}
\caption{Simulation Parameters}
\vspace{-5pt}
\begin{tabular}{cc} \hline
\multirow{2}{*}{Antenna array ($\Delta_d$)} & Non-uniformly weighted\\
& linear array (0.5)\\\hline
Noise figure + implementation loss & 12 dB \\ \hline
Signal bandwidth & 2 GHz \\\hline
\end{tabular}\label{tab:simulation}
\end{small}
\vspace{-15pt}
\end{table}

\subsection{Receive power variations}
\begin{figure}
\centering
\epsfig{file=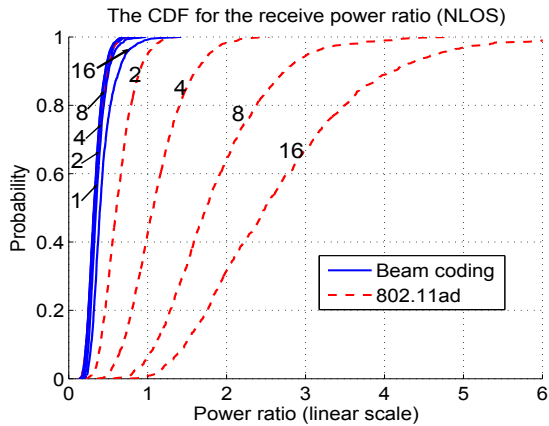,height=2.2in, width=3.2in}
\caption{shows the Cumulative Density Function (CDF) of power ratio for two schemes under NLOS environment. The labels on the curves represent the number of beams (i.e. 1, 2, 4, 8, 16) used in a packet.}
\label{fig:power_var_NLOS}
\end{figure}
\begin{figure}
\centering
\epsfig{file=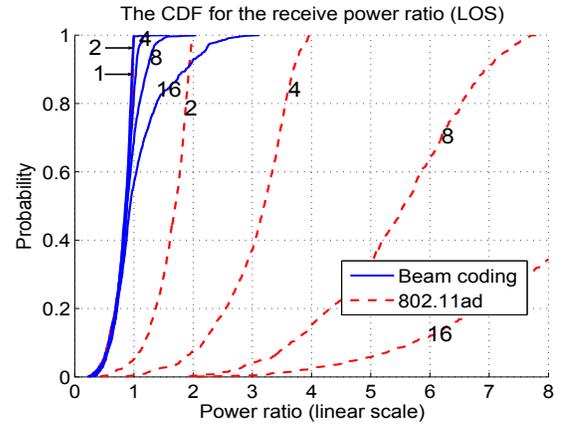,height=2.2in, width=3.2in}
\caption{shows the CDF of power ratio under LOS enviroment.}\label{fig:power_var_LOS}
\end{figure}

The main metric is the power ratio, $\gamma$, which is equal to $\frac{P_{train}}{3\sigma_{prem}}$ where $P_{train}$ is the power of signal in the training section while $\sigma_{prem}$ is the variance of signal in the preamble. The $3\sigma_{prem}$ is used because the AGC gain is to make sure 99.5\% of signals fall within a desired range. The receiver has only one antenna in order to show the worst case scenario of beam coding as it observes more multiple paths if receiving omni-directionally. The transmitter has 16 antennas and varies the number of training beams in a training packet.

The cumulative density function of power ratios in NLOS and LOS environments are shown in Fig. \ref{fig:power_var_NLOS} and \ref{fig:power_var_LOS}. In NLOS, it can be observed that all the training data in beam coding is with $\gamma<1$ while the ratios in the 802.11ad scheme sweep from 0 to 6. In LOS, this effect is exaggerated. The ratio in the 802.11ad scheme can reach 14 for 16 beams in a packet. Though the ratio in beam coding reaches 2.5 in LOS, no extra AGC resetting is required because 95\% of training data is within power ratio 2 and the existence of the dominant path in LOS environment allows a small percentage of power clippings in the training result. Therefore, AGC fields are not needed in the beam coding scheme.

Due to the fact that the beam coding scheme does not need extra AGC resetting nor extra fields for estimating the relative delay of channel taps among various beam directions, beam coding technique can greatly reduce the size of training packets designed in the 802.11ad standard. In the 802.11ad standard, for each beam direction that is being trained, there are 4*320 bits for AGC resetting, 4*640 bits for relative delay estimations and 1024 bits for CE sequences in a training packet while in the beam coding scheme, only 1024 bits for the CE sequence are needed. A saving of 4000 bits per beam direction in the BF training section can be observed by using the proposed beam coding technique.

\subsection{Uniformly weight linear array with phase quantization}
\begin{figure}
\centering
\epsfig{file=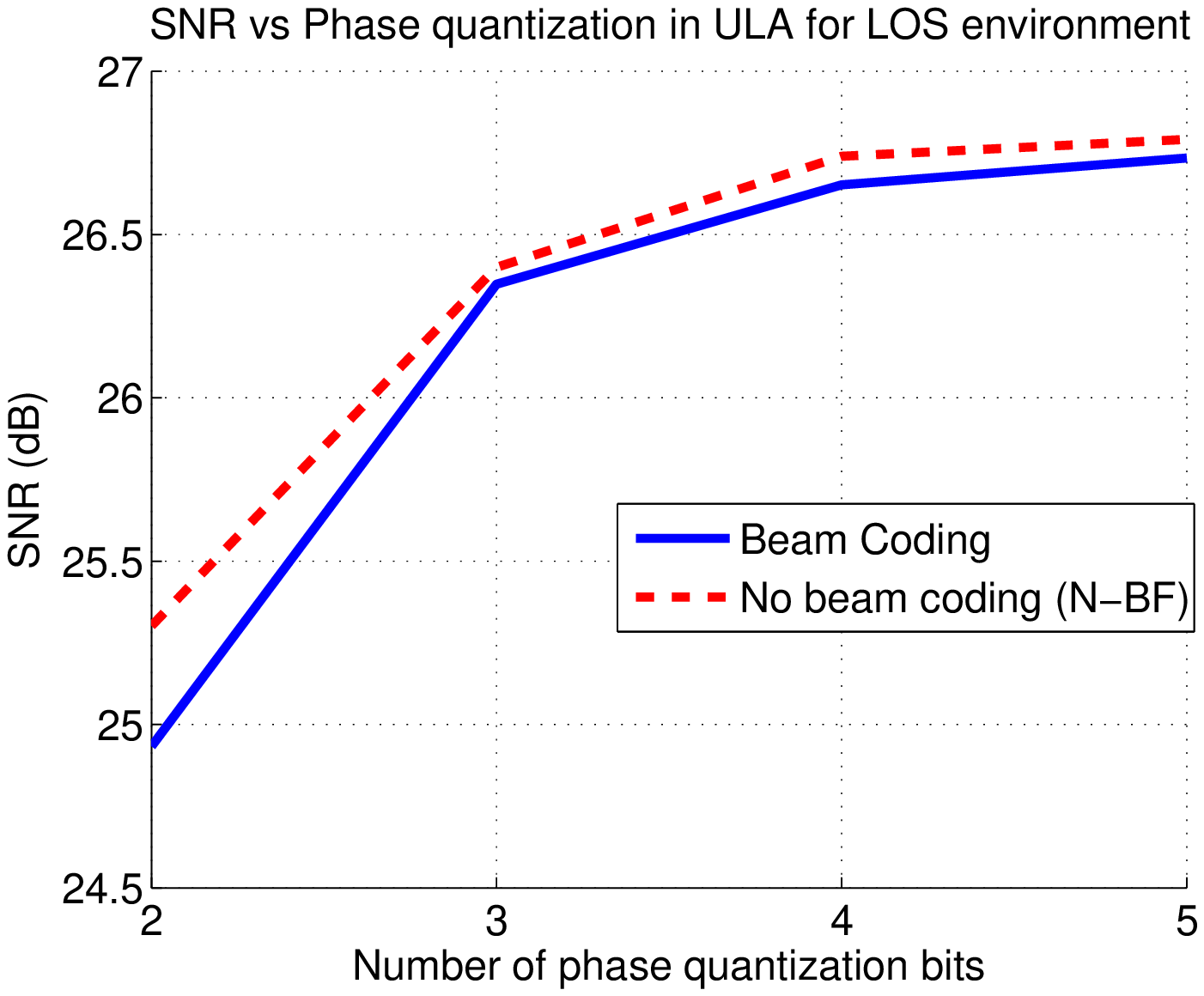,height=2.2in, width=3.2in}
\caption{shows the performance with quantization of phases (LOS)}
\label{fig:phase_quant}
\vspace{+15pt}
\centering
\epsfig{file=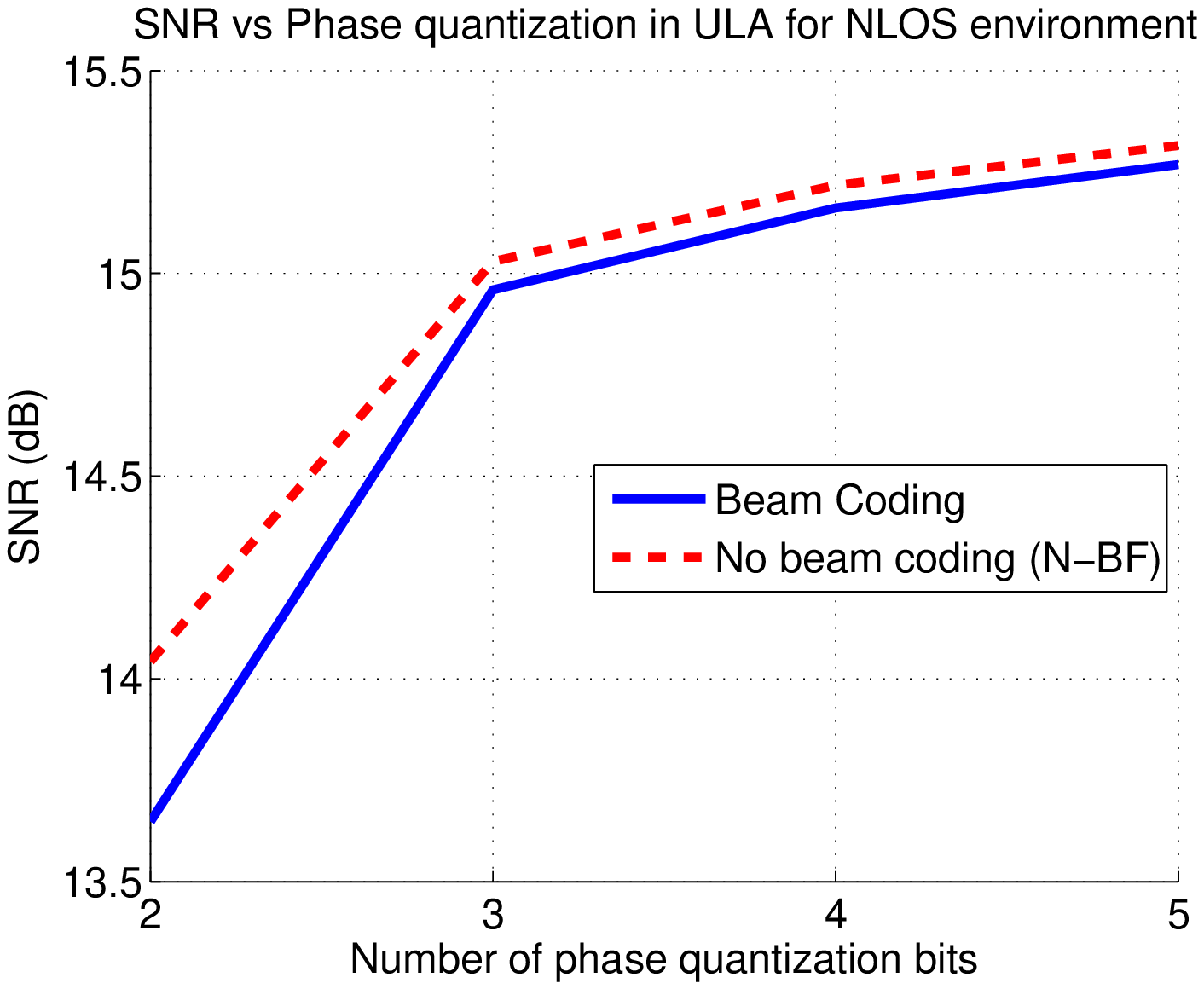,height=2.2in, width=3.2in}
\caption{shows the performance with quantization of phases (NLOS)}\label{fig:phase_quant_NLOS}
\vspace{+5pt}
\centering
 \subfloat[2-bit phase
 quantization]{\epsfig{file=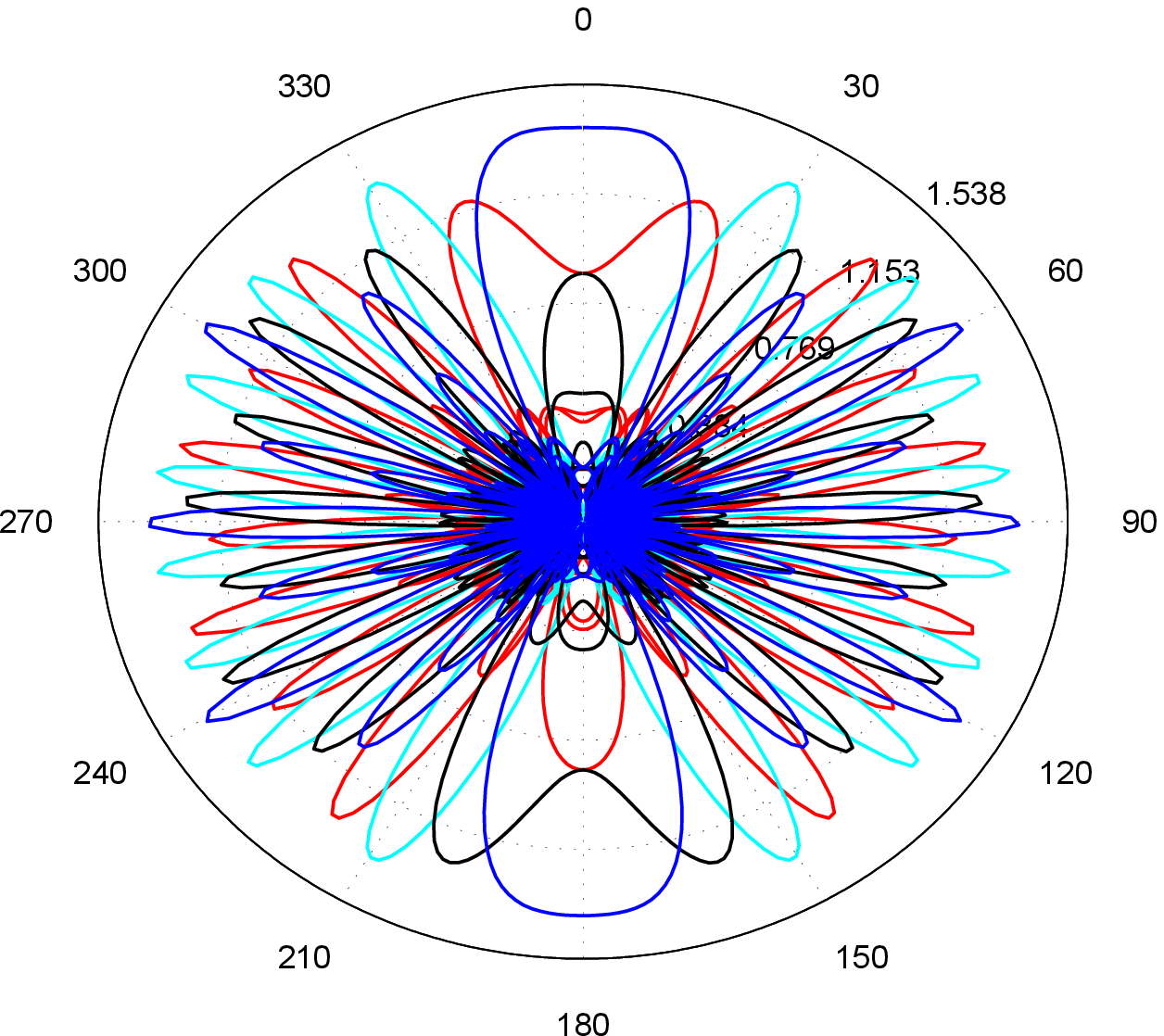,height=1.7in,
 width=1.75in} }
 \subfloat[4-bit phase
 quantization]{\epsfig{file=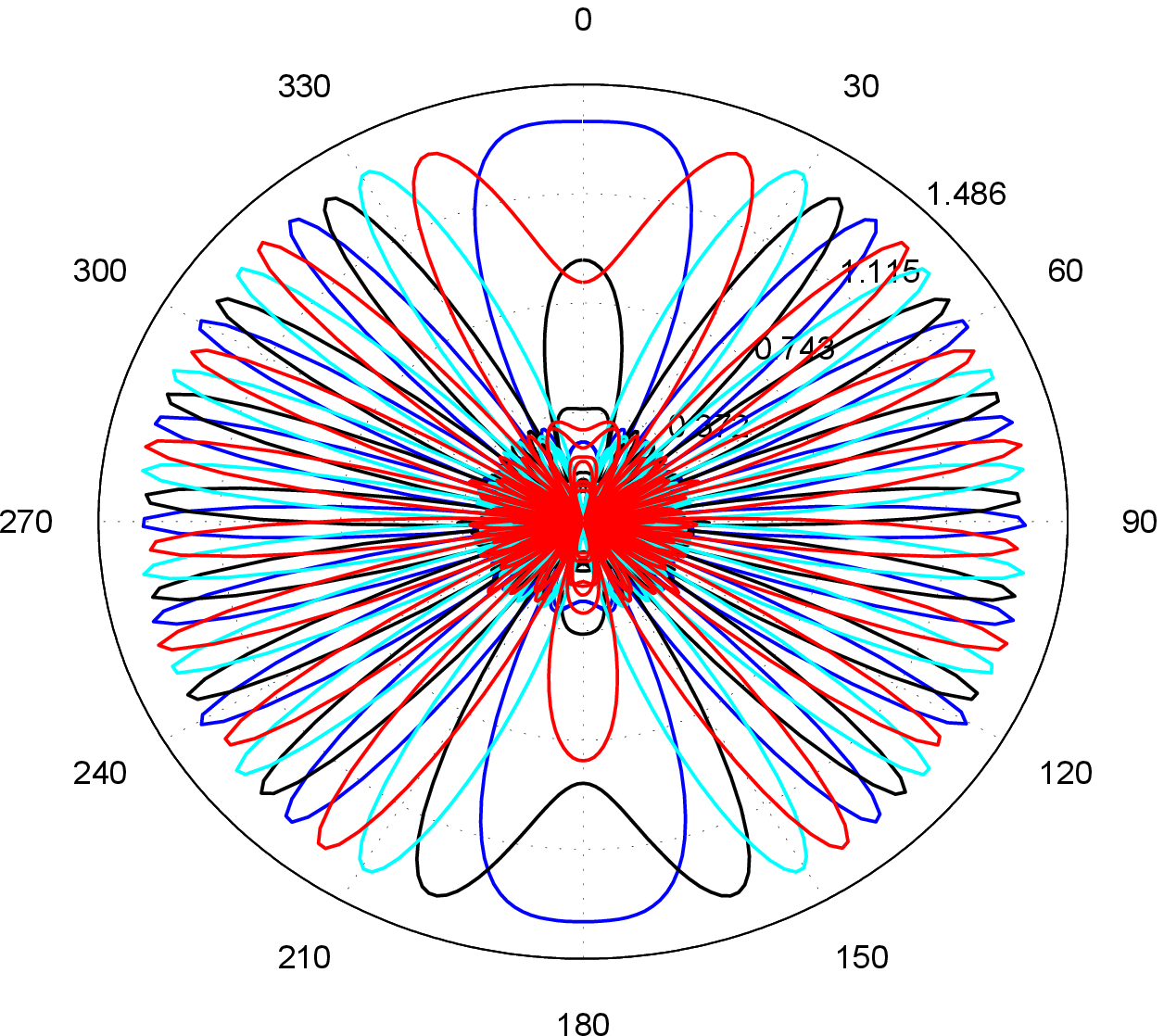,height=1.7in,
 width=1.75in}}
 \caption{shows the beam pattern after beam coding.}\label{fig:polar_plot_2}
 \end{figure}

Phase quantization is commonly found in the mmWave RFIC since the phase shifters in the RFIC are digitally-controlled. In Fig. \ref{fig:phase_quant} and \ref{fig:phase_quant_NLOS}, we compare the exhaustive beam coding scheme with the scheme with No beam coding (N-BF). N-BF scheme is the exhaustive PbP scheme that serves as the upper bound of the performance since it has the best performance in NLOS environment and does not perform multiple-beam steering. The linear scale of SNR in the figure is defined as \emph{SNR}=$2^{\left(\frac{1}{W}\sum_i
\log_2(1+SNR_i)\right)}-1$, where $W$ is the total number of simulation runs and $i$ is the index of a simulation run. In the simulation, transmitter and receiver are having 16 antennas. In each training packet, 16 beam directions are trained at the same time in the beam coding scheme so that it gives the worse possible performance of beam coding due to phase quantization.

Even though beam coding scheme is expected
to give larger side lobes in ULA, its performance approaches that of the optimal exhaustive PbP scheme even with two-bit phase quantization as shown in Fig. \ref{fig:phase_quant}. In NLOS environment, it is worth noting that when the number of phase quantization bits reaches 3, the beam coding scheme has exactly the same performance as the exhaustive PbP scheme. This is explained in Fig. \ref{fig:polar_plot_2}. Almost no distortion in the plots of beam patterns can be seen in the beam coding scheme with 4-bit phase quantization. Though more distortions are observed for 2-bit phase quantization, since its pointing directions are maintained correctly, the performance for beam coding is still very satisfactory even with phase quantization in ULA.

\section{Conclusion}
We present the beam coding scheme to unleash the potentials of the in-packet BF training packet structures for
mmWave system BF training. The scheme provides uniform receive power variations for in-packet training so that no extra AGC resetting is required even though beam directions are varied within a packet. This not only allows in-packet BF training protocols to be executed at every stage of BF training, but also greatly reduces the size of a BF training packet used in the IEEE 802.11ad standard. More importantly, the beam coding scheme does not require variable amplitudes of antenna weights and hence only mild modification of mmwave RF frontend is sufficient to support the beam coding scheme.

\bibliographystyle{IEEEtran}
\bibliography{IEEEabrv,beamCoding_Journal}

\end{document}